\documentclass[useAMS,usenatbib,usegraphicx]{mn2e}
\usepackage{url}
\usepackage{hyperref}
\usepackage{stmaryrd}
\usepackage{supertabular}

\def\aj{AJ}%
\def\apj{ApJ}%
\def\apjl{ApJ}%
\def\apjs{ApJS}%
\def\aap{A\&A}%
\def\aaps{A\&AS}%
\def\mnras{MNRAS}%
\def\nat{Nature}%
\def\physrep{Phys.~Rep.}%
\def\nsamp{141 }

\def\hi{H\,{\sc i}}
\def\m20{$M_{20}$}
\begin{document}

\title[Lopsidedness and Interaction in \hi \ Maps]{Quantified \hi \ Morphology II : \\ Lopsidedness and Interaction in WHISP Column Density Maps}

\author[B.W. Holwerda et al.]{B. W. Holwerda$^{1,2}$\thanks{E-mail:
benne.holwerda@esa.int}, N. Pirzkal,$^{3}$ W.J.G. de Blok,$^{2}$ A. Bouchard,$^{4}$ S-L. Blyth,$^{2}$ 
\newauthor
K. J. van der Heyden$^{2}$ and E.C. Elson$^{2}$\\
$^{1}$ European Space Agency, ESTEC, Keplerlaan 1, 2200 AG, Noordwijk, the Netherlands\\
$^{2}$ Astrophysics, Cosmology and Gravity Centre ($ACGC$), \\
Astronomy Department, University of Cape Town, Private Bag X3, 7700 Rondebosch, Republic of South Africa\\
$^{3}$ Space Telescope Science Institute, 3700 San Martin Drive, Baltimore, MD 21218, USA\\
$^{4}$ Department of Physics, Rutherford Physics Building, McGill University, 3600 University Street, Montreal, Quebec, H3A 2T8, Canada}

\date{Accepted -- Received -- ; in original form --}

\pagerange{\pageref{firstpage}--\pageref{lastpage}} \pubyear{2010}

\maketitle

\label{firstpage}

\begin{abstract}

Lopsidedness of the gaseous disk of spiral galaxies is a common phenomenon in disk morphology, profile and kinematics. 
Simultaneously, the asymmetry of a galaxy's stellar disk, in combination with other morphological parameters, has seen 
extensive use as an indication of recent merger or interaction in galaxy samples. Quantified morphology of stellar spiral 
disks is one avenue to determine the merger rate over much of the age of the Universe. 
In this paper, we measure the quantitative morphology parameters for the \hi \ column density maps from the Westerbork 
observations of neutral Hydrogen in Irregular and SPiral galaxies (WHISP). These are Concentration, Asymmetry, 
Smoothness, Gini, $M_{20}$, and one addition of our own, the Gini parameter of the second order moment ($G_M$). 
Our aim is to determine if lopsided or interacting disks can be identified with these parameters. Our sample of \nsamp \ \hi \ 
maps have all previous classifications on their lopsidedness and interaction.

We find that the Asymmetry, $M_{20}$ and our new $G_M$ parameter correlate only weakly with the previous morphological 
lopsidedness quantification. These three parameters may be used to compute a probability that an \hi \ disk is morphologically 
lopsided but not unequivocally to determine it. However, we do find that that the question whether or not an \hi \ disk is interacting 
can be settled well using morphological parameters. Parameter cuts from the literature do not translate from ultraviolet to \hi \ 
directly but new selection criteria using combinations of Asymmetry and $M_{20}$ or Concentration and $M_{20}$, work very well. 

We suggest that future all-sky \hi \ surveys may use these parameters of the column density maps to determine the merger fraction 
and hence rate in the local Universe with a high degree of accuracy.

\end{abstract}


\section{\label{s:intro}Introduction}


In the study of the 21 cm emission of atomic hydrogen (\hi) from nearby galaxies it was noted early on that many appear to be not symmetric. This phenomenon was termed ``lopsidedness'' of the \hi \ morphology \citep{Baldwin80}. \cite{Richter94} find that half of galaxy disks are lopsided.
A similar deviation from the axi-symmetry in the stellar disks of galaxies was noted by \cite{Rix95} and \cite{Zaritsky97}. 
The \hi \ line profile of half the population of galaxies, also shows a clear deviation from symmetry on either side of the systemic velocity \citep{Haynes98,Matthews98}. \cite{Swaters99} report a third kind of lopsidedness, a deviation from axi-symmetry in the position-velocity diagram, termed kinematic lopsidedness. The fraction of lopsided galaxies may depend on environment; e.g., the Eridanus Galaxy group counts twice as many lopsided galaxies as the field \citep{Angiras06}, but the Ursa Major group is similar to the field \citep{Angiras07}. Lopsidedness seems to be the strongest in the outer regions of a disk \citep{Jog99}.

Disk lopsidedness may be the product of tidal interactions \citep{Jog97}, minor mergers \citep{Zaritsky97}, asymmetric accretion of fresh gas from the cosmic web \citep{Bournaud05}, intergalactic gas ram-pressure \citep{Mapelli09}, or an offset between the disk and dark matter halo \citep[``disk sloshing'',][]{Levine98, Noordermeer01}. Alternatively, the phenomenon may be attributed equally to most of these causes \citep{Mapelli09}. In the case of a stellar disk, lopsidedness may have an internal cause, such as a dynamical instability \citep[e.g.,][]{Lovelace99, Dury08}.

Initially, the main way to identify lopsidedness has been a visual inspection of the \hi \ column density map, line profile or velocity field of a galaxy \citep{Richter94, Haynes98, Matthews98, Swaters02, Noordermeer05}. In addition, lopsidedness can be quantified using a Fourier decomposition of either the stellar image \citep{Zaritsky97, Bournaud05}, the \hi \ column density map \citep{Angiras06, Angiras07}, or velocity field \citep{Schoenmakers97,Trachternach08}. However, the Fourier analysis has only been performed on small samples of \hi \ observations or larger samples of optical ones. We refer the reader to the in-depth review by \cite{Jog09} on the lopsidedness phenomenon.

In parallel with the line of investigation into galaxy lopsidedness, a considerable observational effort has gone into morphological tracers of interaction over cosmological times. These studies use certain quantifiable morphological parameters of restframe-ultraviolet images of distant and nearby galaxies to estimate the merger rate of galaxies \citep{Abraham94,Conselice00a, Lotz04}. Two sets of parameterisations have emerged, the Concentration-Asymmetry-Smoothness by \cite{CAS} and the Gini-$M_{20}$ by \cite{Lotz04}.
These parameterisations of galaxy morphology have now been applied on every deep multi-wavelength Hubble field to determine merger rates; in the Hubble Ultra Deep Field by \cite{Pirzkal06}, GOODS by \cite{Bundy05} and \cite{Ravindranath06}, COSMOS by \cite{Scarlata07} and \cite{Conselice09b}, GEMS by \cite{Jogee09}, and the extended Groth strip by \cite{Lotz08b} and \cite{Conselice08b}, as well as local reference samples \citep{CAS, Lotz04, Bendo07,Munoz-Mateos09}.
The different parameterisations of galaxy appearance are sensitive to different stages of an interaction and different interaction types \citep{Lotz08, Conselice09c,Lotz10a,Lotz10b} but are very successful in estimating the galaxy merger fraction and rate over much of the age of the Universe. 

Yet, to date, these studies have been constrained mostly to restframe ultraviolet and optical because these are the wavelengths where interaction-induced star-formation produces high-surface brightness features in galaxies with clearly disturbed morphology at wavelengths where the Hubble Space Telescope can reasonably observe them. 

In the previous papers in this series \citep{Holwerdapra09, Holwerda10a,Holwerda10b}, we have shown that a description of the \hi \ morphology using these parameters is as sensitive, if not better than, any of the star-formation dominated wavelengths to the effects of interactions. Hence, the future Square Kilometer Array \citep[SKA;][]{ska} and its precursor radio telescopes, South Africa's Karoo Array Telescope \citep[MeerKAT;][]{MeerKAT,meerkat1,meerkat2}, and the Australian SKA Pathfinder \citep[ASKAP;][]{askap2, askap1, ASKAP, askap3,askap4}, provide an opportunity to explore the lopsidedness phenomenon as well as interactions using 21 cm line emission (\hi) of thousands of galaxies. However, to do so, automated parameterisations of the \hi \ maps are needed and the relation between the lopsidedness phenomena and the above parameter space will need to be explored. If, for instance a parameter space can be identified for interacting galaxies, a merger fraction for a given cosmic volume can be measured. Combined with an estimate of the timescale a merger spends in this parameter space \citep{Holwerda10d}, one can then estimate the merger rate \citep{Holwerda10e}. Our ultimate goal is to simplify the selection of subsamples in the upcoming large \hi \ surveys using existing morphological parameters.

In this paper, we compare the CAS and Gini-$M_{20}$ parameters as determined in the \hi \ column density maps of \nsamp galaxies, to the lopsidedness qualification and interaction determinations from \cite{Swaters02} and \cite{Noordermeer05}. In \S \ref{s:def}, we define the two main concepts, in \S \ref{s:morph} we briefly discuss the morphological parameters and present a new additional parameter. In \S \ref{s:whisp}, we describe the radio data used. Our results are presented in \S \ref{s:results} with our conclusions in \S \ref{s:concl} and a brief outlook in \S \ref{s:fut}.


\section{Lopsidedness and Asymmetry}
\label{s:def}

In the following discussion, we find it useful to define the terms lopsidedness and asymmetry as they are often used interchangeably and the difference is subtle. \\
{\em Lopsidedness} is a comparison of {\em axi}-symmetry, the level of symmetry of a galaxy image, line profile or velocity field when mirrored over an axis (minor or major) \citep{Baldwin80,Swaters99}. Quantified definitions of lopsidedness are when a disk displays an $m=1$ global spatial offset (m is the azimuthal wavenumber in a spatial Fourier decomposition) or the $cos(\phi)$ distribution ($\phi$ is the azimuthal angle) is non-axisymmetric. In the case of the WHISP sample, there are qualitative estimates of a disk's lopsidedness available from \cite{Swaters02} and \cite{Noordermeer05}.\\
{\em Asymmetry}, as defined by \cite{Abraham94} and \cite{CAS}, is the {\em point}-symmetry of an object; the level of symmetry when the object is rotated $180^\circ$ around its centre.

\section{Morphological Parameters}
\label{s:morph}

The morphological parameters we compute over the \hi \ column density maps are Concentration, Asymmetry and Smoothness from \cite{CAS}, $M_{20}$ and Gini from \cite{Lotz04}, and a single addition of our own; the Gini parameter of the second order moment of the light, $G_M$. We describe our implementation of the existing parameters in \cite{Holwerda10a,Holwerda10b} and below. The relevant input parameters are the central position of the galaxy ($x_c$, $y_c$), and a definition of the area over which these parameters are computed. The Gini parameter only requires the definition of the area and not the central position, making it less sensitive to input error. We obtained uncertainty estimates from a Monte-Carlo run, varying the central position of each galaxy and a separate run randomly redistributing the pixel values in the galaxy-area. 

\subsection{CAS}
\label{ss:cas}

CAS refers to the now commonly used Concentration-Asymmetry-Smoothness space \citep{CAS} for morphological analysis of distant galaxies. Concentration of the light, symmetry around the centre and smoothness as an indication of substructure.

Concentration is defined by \cite{Bershady00} as:
\begin{equation}
C = 5 ~ \log (r_{80} /  r_{20})
\label{eq:c}
\end{equation}
\noindent with $r_{f}$ as the radius containing percentage $f$ of the light of the galaxy \citep[see definitions of $r_f$ in][]{se,seman}.

The asymmetry is defined as the level of {\em point}-, (or rotational-) symmetry around the centre of the galaxy \citep{Abraham94,CAS}:
\begin{equation}
A = {\Sigma_{i,j} | I(i,j) - I_{180}(i,j) |  \over \Sigma_{i,j} | I(i,j) |  },
\label{eq:a}
\end{equation}
\noindent where $I(i,j)$ is the value of the pixel at the position $i,j$ in the image, and $I_{180}(i,j)$ is the pixel at position $[i,j]$ in the galaxy's image, after it was rotated $180^\circ$ around the centre of the galaxy.

Inspired by the ``unsharp masking" technique \citep{Malin78b}, Smoothness is defined by \cite{Takamiya99} and \cite{CAS} as:
\begin{equation}
S = {\Sigma_{i,j} | I(i,j) - I_{S}(i,j) | \over \Sigma_{i,j} | I(i,j) | }
\label{eq:s}
\end{equation}
\noindent where $I_{S}(i,j)$ is the same pixel in a smoothed image. What type of smoothing is used has changed over the years. We chose a fixed 5" Gaussian smoothing kernel for simplicity.

\subsection{Gini and $M_{20}$}
\label{ss:gm20}

\cite{Abraham03} and \cite{Lotz04} introduce the Gini parameter to quantify the distribution of flux over the pixels in an image.
They use the following definition:
\begin{equation}
G = {1\over \bar{I} n (n-1)} \Sigma_i (2i - n - 1) I_i ,
\label{eq:g}
\end{equation}
\noindent $I_i$ is the value of pixel i in an ordered list of the pixels, $n$ is the number of pixels in the image, and $\bar{I}$ is the mean pixel value in the image. 
We chose this definition as it is the computationally least expensive. The Gini parameter is an indication of equality in a distribution \citep[initially an economic indicator][]{Gini12,Yitzhaki91}, with G=0 the perfect equality (all pixels have the same intensity) and G=1 perfect inequality (all the intensity is in a single pixel). Its behaviour is therefore in between that of a structural measure and concentration. 

\cite{Lotz04} also introduced a new way to parameterize the extent of the light in a galaxy image. They define the spatial second order moment as the product of the intensity with the square of the projected distance to the centre of the galaxy. This gives more weight to emission further out in the disk. It is sensitive to substructures such as spiral arms and star-forming regions but insensitive if these are distributed symmetrically or not.

The second order moment of a pixel $i$ is defined as:
\begin{equation}
M_i = I_i \times [(x-x_c)^2 + (y-y_c)^2 ],
\label{eq:Mi}
\end{equation}
where $[x, y]$ is the position of a pixel with intensity value $I_i$ in the image and $[x_c, y_c]$ is the central pixel position of the galaxy in the \hi \ surface density map.

The total second order moment of the image is given by:
\begin{equation}
M_{tot} = \Sigma_i M_i = \Sigma I_i [(x_i - x_c)^2 + (y_i - y_c)^2].
\label{eq:mtot}
\end{equation}

\cite{Lotz04} use the relative contribution of the brightest 20\% of the pixels to the second order moment as a measure of disturbance of a galaxy:
\begin{equation}
M_{20} = \ log \left( {\Sigma_i M_i  \over  M_{tot}}\right), ~ {\rm for} ~ \Sigma_i I_i < 0.2 I_{tot}. \\
\label{eq:m20}
\end{equation}
The $M_{20}$ parameter is sensitive to bright regions in the outskirts of disks and thus higher values can be expected in galaxy images (in the optical and UV) 
with star-forming outer regions as well as those images of strongly interacting disks.

\subsection{Gini of the second order moment ($G_M$)}
\label{ss:gm}

Instead of using the intensity of pixels, we can define a Gini parameter for the second order moment of each pixel by substituting $M_i$ (equation \ref{eq:Mi}) for $I_i$ in equation \ref{eq:g}:
\begin{equation}
G_M = {1\over \bar{M} n (n-1)} \Sigma_i (2i - n - 1) M_i ,
\label{eq:gm}
\end{equation}

This is our contribution to the parameter space to provide an additional handle to characterise lopsidedness and interaction level. Our reasoning was that the Gini parameter has the added benefit of using the combined shape of the flux distribution curve (all the information in the image), rather than just a fraction. In \cite{Holwerda10b}, we found hints that $M_{20}$ may be not sensitive enough to interaction signature while Asymmetry is sensitive to other effects as well. A similar conclusion was reached by \cite{Lotz08a}, hence $G_M$ is an attempt to define a single parameter to detect interaction using all the information on the second order moment.

\section{WHISP data}
\label{s:whisp}

The data we use are the \hi \ column density maps from the Westerbork \hi \ Survey of Irregular and SPiral galaxies \citep[WHISP,][]{whisp,whisp2,Swaters02,Swaters02b,Noordermeer05}.  WHISP is a survey of the neutral hydrogen component in spiral and irregular galaxies with the Westerbork Synthesis Radio Telescope (WSRT). It has mapped the distribution and velocity structure of \hi \ in several hundreds of nearby galaxies, increasing the number of \hi \ observations of galaxies by an order of magnitude. The WHISP project provides a uniform database of datacubes, zeroth-order and velocity maps. Its focus has been on the structure of the dark matter halo as a function of Hubble type, the Tully-Fisher relation and the dark matter content of dwarf galaxies. 

The WHISP observation targets were selected from the Uppsala General Catalogue of Galaxies \citep{Nilson73}, with blue major diameters $> 2\farcm0$, declination (B1950) $\delta ~ > ~ 20^\circ$ and flux densities at 21-cm larger than 100 mJy, later lowered to 20 mJy Observation times were typically 12 hours of integration. The galaxies satisfying these selection criteria generally have redshifts less than 20000 km/s ($z<0.07$). A further prerequisite was that either \cite{Swaters02} or \cite{Noordermeer05} classified both the level of the galaxy's lopsidedness and whether or not it is interacting. 

The WHISP data was retrieved from the ``Westerbork on the web'' project at ASTRON (\url{http://www.astron.nl/wow/}).  We use the column density maps with the highest resolution available ($\sim$12" x 12"/sin($\delta$)). The positions and basic \hi \ information (masses and diameters etc.) are from \cite{Swaters02} and \cite{Noordermeer05cat,Noordermeer05}.
We used the central position ($x_c,y_c$) as input for the parameters and the radius of the \hi \ disk ($R_{HI}$) to cut out a stamp of the disk before computation (a stamp was set at $7 \times R_{HI}$). The computed morphological parameters are in Tables A1 and A2 in Appendix A ({\em electronic edition only}).


\section{Results}
\label{s:results}

The samples from \cite{Swaters02} and \cite{Noordermeer05} both have visual classifications of a galaxy's lopsidedness and whether or not it is interacting.
Lopsidedness was determined for each galaxy's morphology in the \hi \ column density map, its profile and velocity map. These are the morphological, profile and kinematic lopsidedness respectively. These classification were done by a single observer by visual inspection and hence carry some risk of observer bias.
The classifications by \cite{Swaters02} for lopsidedness are: not-, weak-, and strong lopsidedness.  
The lopsidedness classification by \cite{Noordermeer05} is a little more nuanced with no, mildy-, moderately-, and severely lopsided. 
Both authors classify morphological, profile as well as kinematic lopsidedness with their respective qualifiers. 
To unify the two classification schemes, we re-assigned the \cite{Swaters02} classifications to lopsidedness categories of the \cite{Noordermeer05} classification: 
weak is equivalent to mild and strong to severely. In the following section we use the Noordermeer et al. lopsidedness scale.

\cite{Noordermeer05} also gives an estimate on whether the galaxy is interacting and \cite{Swaters02} lists the five galaxies in their sample of 74 dwarfs that are 
in an active interaction. This fraction (5/74) might be an under-estimate and may not account for galaxies that are only mildly interacting.

In the following section (\ref{ss:lob}), we compare the visual classifications of lopsidedness to our morphological parameters to determine if lopsidedness can be quantified with our morphological parameters. In the next section (\ref{ss:int}), we explore the distribution of interacting galaxies in our parameter space. 
These visual classifications are subject to possible observer bias but the aim here is to identify the parts of the morphological parameter space described in section \ref{s:morph} that hold the majority of lopsided or interacting galaxies. 
Appendix A ({\em electronic edition only}) lists the morphological parameters of all the galaxies in the Swaters et al. (2002) and Noordermeer et al. (2005) sample in Tables A1 and A2 respectively. 

\begin{figure*}
\centering
\includegraphics[width=\textwidth]{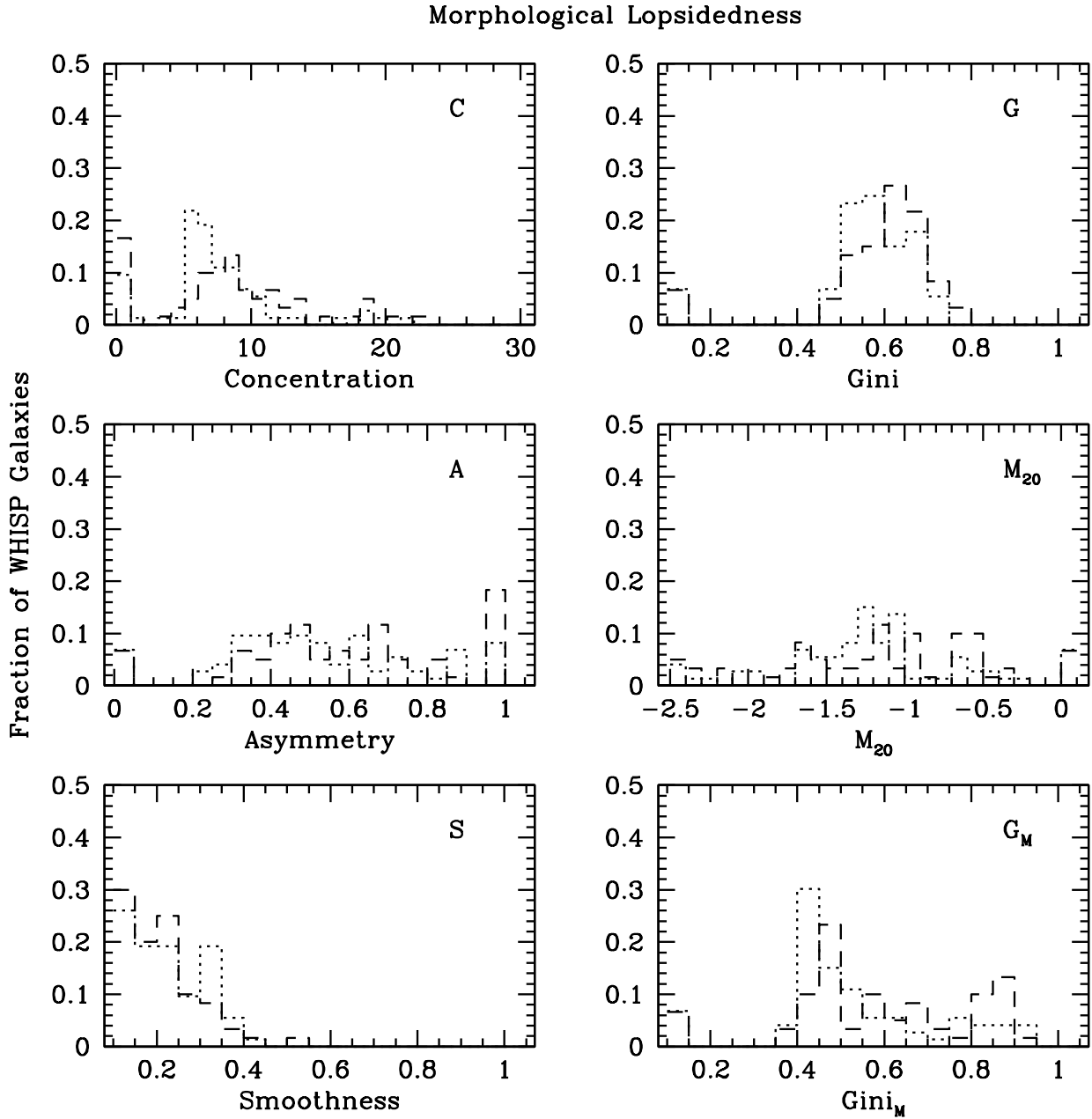}
\caption{\label{f:2d:hist:mlob} The normalized distribution of parameters of the combined samples from \protect\cite{Swaters02} and \protect\cite{Noordermeer05} for morphologically lopsided (dashed) and non-lopsided (dotted) histograms. There are small shifts in the distributions of Concentration, Gini and to a lesser extent $M_{20}$ and $G_M$  but no clear separation between the lopsided and non-lopsided galaxies.}
\end{figure*}

\begin{figure*}
\centering
\includegraphics[width=\textwidth]{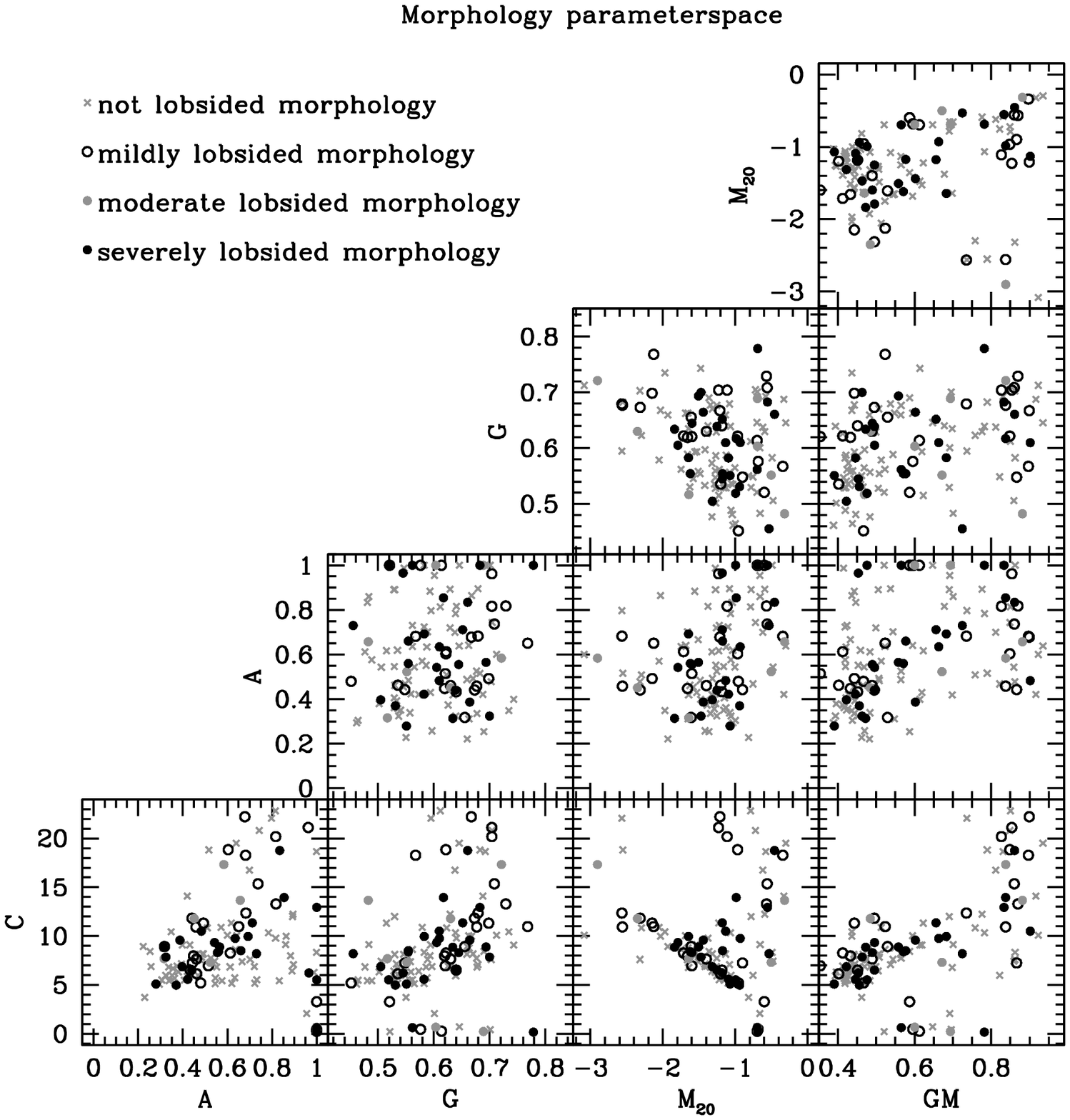}
\caption{\label{f:2d:spr:lop} The distribution of parameters of the combined samples from \protect\cite{Swaters02} and \protect\cite{Noordermeer05} for the different morphologically lopsided classifications (no, mild, moderate, and severe). There seems to be no clear preference of a lopsidedness classification for any part of the parameter space.}
\end{figure*}

\subsection{Lopsidedness}
\label{ss:lob}

Figure \ref{f:2d:hist:mlob} shows the histograms of the above six parameters (C, A, S, G, $M_{20}$, and $G_M$) for lopsided (any strength) and non-lopsided galaxy morphology according to either \cite{Swaters02} or \cite{Noordermeer05}. 
Based on the definition of lopsidedness, we expected the Asymmetry, $M_{20}$, and possibly the $G_M$  parameters to show a difference between the two populations. We observe a difference in the median in these parameters with the lopsided galaxies showing higher values for Asymmetry and $G_M$ and lower values for $M_{20}$ (Figure \ref{f:2d:hist:mlob}). The Gini parameter shows a shift in the median value for the lopsided galaxies as well. 

Figure \ref{f:2d:spr:lop} shows the parameter space of C, A, G, $M_{20}$, and $G_M$ with the different lopsidedness classifications (no, mild, moderate and severely). As mentioned above, weak and strong according to \cite{Swaters02} are plotted as mild and severely respectively. There is no clear part of parameter space where one could identify only, for instance, the severely (strongly) lopsided galaxies. 

While the distributions of morphological values are different between lopsided and not lopsided galaxies, there is no clear cut in morphological parameters to discern between the two or separate out weakly and strongly lopsided galaxies. As it stands, the distributions in Figure \ref{f:2d:hist:mlob} could be used to compute a probability that a galaxy is lopsided, but this would still have to be followed up with a visual inspection like those in \cite{Swaters02} and \cite{Noordermeer05} or a Fourier decomposition such as the ones in \cite{Zaritsky97, Bournaud05, Angiras06, Angiras07, van-Eymeren11b}. 

To illustrate further, Figures B1 and B2 in 
Appendix B ({\em electronic version}) show some typical \hi \ maps from the WHISP sample for the minimum, mean and maximum values of Asymmetry and Gini for all four lopsidedness categories; none, weak, moderate and strong lopsidedness. In our view, these images illustrate how, for instance, Asymmetry and qualitative lopsidedness do not measure the same thing. A galaxy can be strongly asymmetric with a lot of flux in a spiral arm offset from the centre of the galaxy, while at the same time, the outer contour may appear much like a ordinary disk. Conversely a strung-out galaxy may appear very lopsided at the lowest \hi \ flux levels, but if there is little flux in the outermost part, and a strong, symmetric disk (with a ring for instance), this may not show in any of the morphological parameters. Our parameters are flux-weighted by design but the qualitative visual classification of lopsidedness by the previous authors may not be.

To verify if the shape of the outer contour alone is a better indication of lopsidedness, we compared the morphological parameters for the images with uniform weighting (pixel-values set to $I_i = 1$). The Gini parameter, in this case, is of no use as this image is perfectly equal (G=0). The Asymmetry, $M_{20}$ and $G_M$ parameters show less change in distribution between the lopsided and non-lopsided populations.
 
Alternatively, in order to parameterize morphological lopsidedness in similar terms as the morphological parameters presented, one could redefine Asymmetry using a specific axis (requiring the additional input of a Position Angle).  For example, \cite{Baldwin80} chose the east-west axis and \cite{Richter94} chose the systemic velocity axis. We compared fluxes from either side of the centre of these galaxies for the x and y-axes of the maps ($A_x$ and $A_y$, Tables A1 and A2 in Appendix A, {\em electronic edition}) and found no relation with the lopsidedness qualifier from either \cite{Swaters02} or \cite{Noordermeer05} (see Table \ref{t:swaters} and \ref{t:noordermeer}). Lopsidedness cannot easily be quantified using the above common morphological parameters or simple variations thereof. 
At best, the histograms in Figure  \ref{f:2d:hist:mlob} can be used to assign a probability of morphological lopsidedness. 
In light of the fact that the above morphological parameters were developed to discern spirals from ellipticals in the optical, their insensitivity to the lopsidedness of spiral \hi \ disks is a indication of the extent that they can be used to classify morphological sub-types.
Therefore, a Fourier decomposition \citep[similar to][]{Zaritsky97, Bournaud05, Angiras06, Angiras07} is still needed to classify and quantify the lopsidedness of galaxies in future large \hi \ surveys, such as WALLABY on ASKAP or a northern sky \hi \ survey with APERTIF on WSRT or the MHONGOOSE\footnote{MeerKAT HI Observations of Nearby Galactic Objects: Observing Southern Emitters} nearby galaxy survey with the MeerKAT radio telescope. 

\begin{table*}
\caption{The number of galaxies selected as interacting in the two WHISP subsets; the number and fraction of the sample, and number of individual galaxy in agreement with the visual classification by either author. The first three criteria are for the CAS and Gini/$M_{20}$ used in the literature. We defined criteria 3, 4 and 5 for \hi \  morphology. The last three are various combination of our \hi \ criteria. The criteria from the literature (1, 2 and 3) are for optical morphology and overselect \hi \ disks compared to the visual classification. Of the \hi \ criteria, (5) and (6) work well, with the latter agreeing with the visual classification in the case of individual galaxies. }
\begin{center}
\begin{tabular}{l l l | l l }
								& Noordermeer		& 			& Swaters 	& 	\\	
								& (68 galaxies)		&			& (73 galaxies) & 	\\
\hline
Selection Criterion 					& Nr. (fraction)		& Individual	& Nr. (fraction)	& Individual	\\
								& 				& agreement	& 			& agreement \\								
\hline
\hline
Visual classification					& 27	(39\%)		& -			& 5 (7\%) 		& -	\\
\hline
(1)	$A > 0.38$ 					& 55 (80\%)		& 23			& 53 (78\%)	& 4\\
(2)	$G > -0.133 \times M_{20} + 0.384$ 	& 51 (75\%)		& 20			& 43 (63\%)	& 4\\
(3)	$G > -0.4 \times A + 0.66$ 		& 61 (81\%)		& 23			& 65 (96\%)	& 5\\
\hline
(4)	$G_M > 0.6$		 			& 39 (57\%)		& 21			& 13 (19\%)	& 4\\ 
(5)	$A < -0.2 \times M_{20} + 0.25$	& 22 (32\%)		& 2			& 36 (52\%)	& 1\\
(6)	$C > -5 \times M_{20} + 3$		& 23 (33\%)		& 11			& 8   (11\%)	& 4\\
\hline
(4) \& (5)							& 6 (8\%)			& 1			& 5(7\%)		& 0\\
(5) \& (6)							& 0 (0\%)			& 0			& 1(1\%)		& 0\\
(4) \& (6)							& 23 (33\%)		& 11			& 8(11\%)		& 4\\

\hline
\end{tabular}
\end{center}
\label{t:frac}
\end{table*}%

\begin{figure*}
\centering
\includegraphics[width=\textwidth]{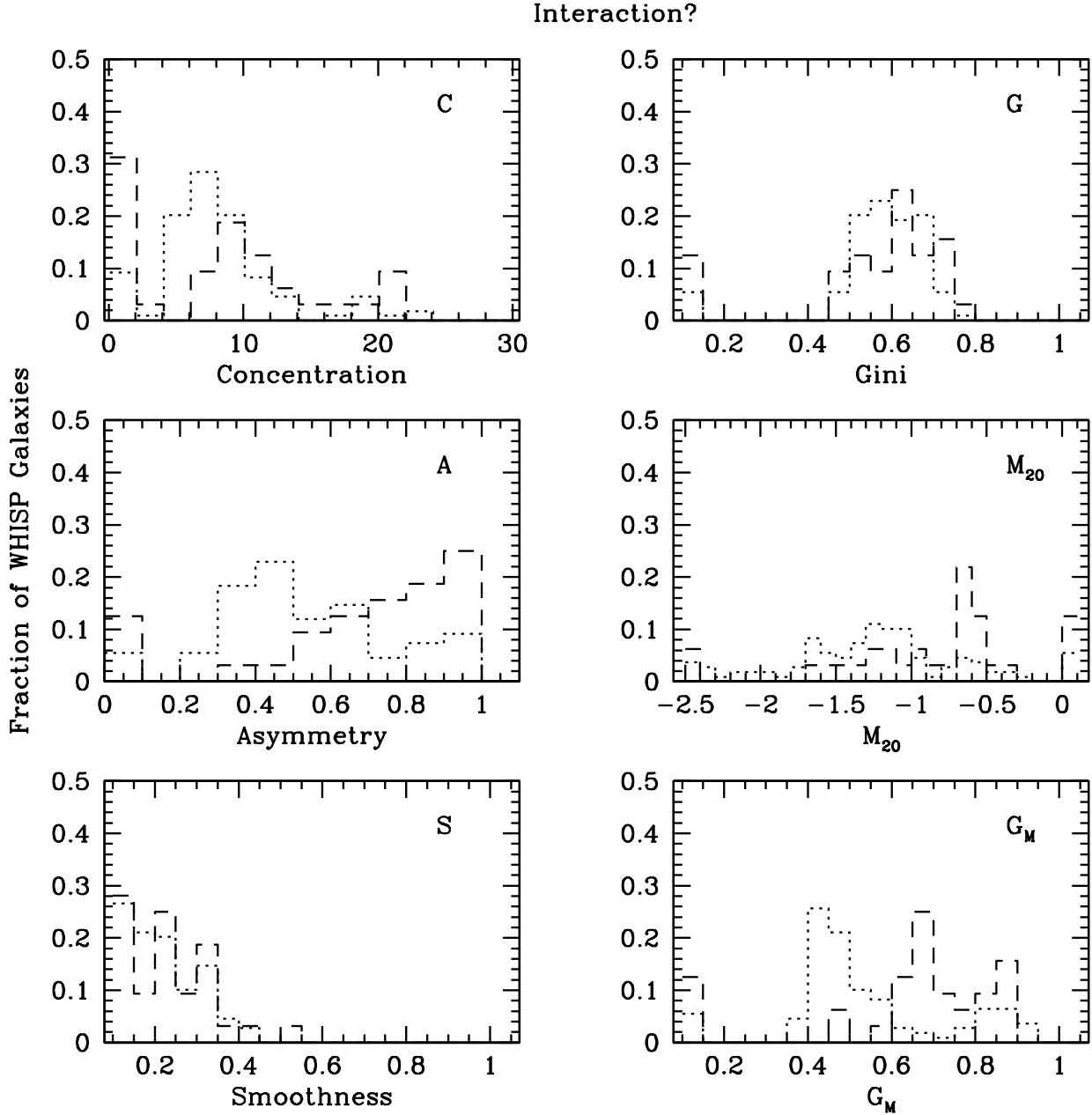}
\caption{\label{f:2d:hist:int} The distribution of parameters of the combined samples from  \protect\cite{Swaters02} and \protect\cite{Noordermeer05} for interacting (dashed) and non-interacting (dotted histograms). There are clear separations in the distribution of Asymmetry, $M_{20}$ and $G_M$ values between the two populations of galaxies. Values in Tables A1 and A2 {\em in the electronic version of the manuscript}. }
\end{figure*}

\begin{figure*}
\centering
\includegraphics[width=\textwidth]{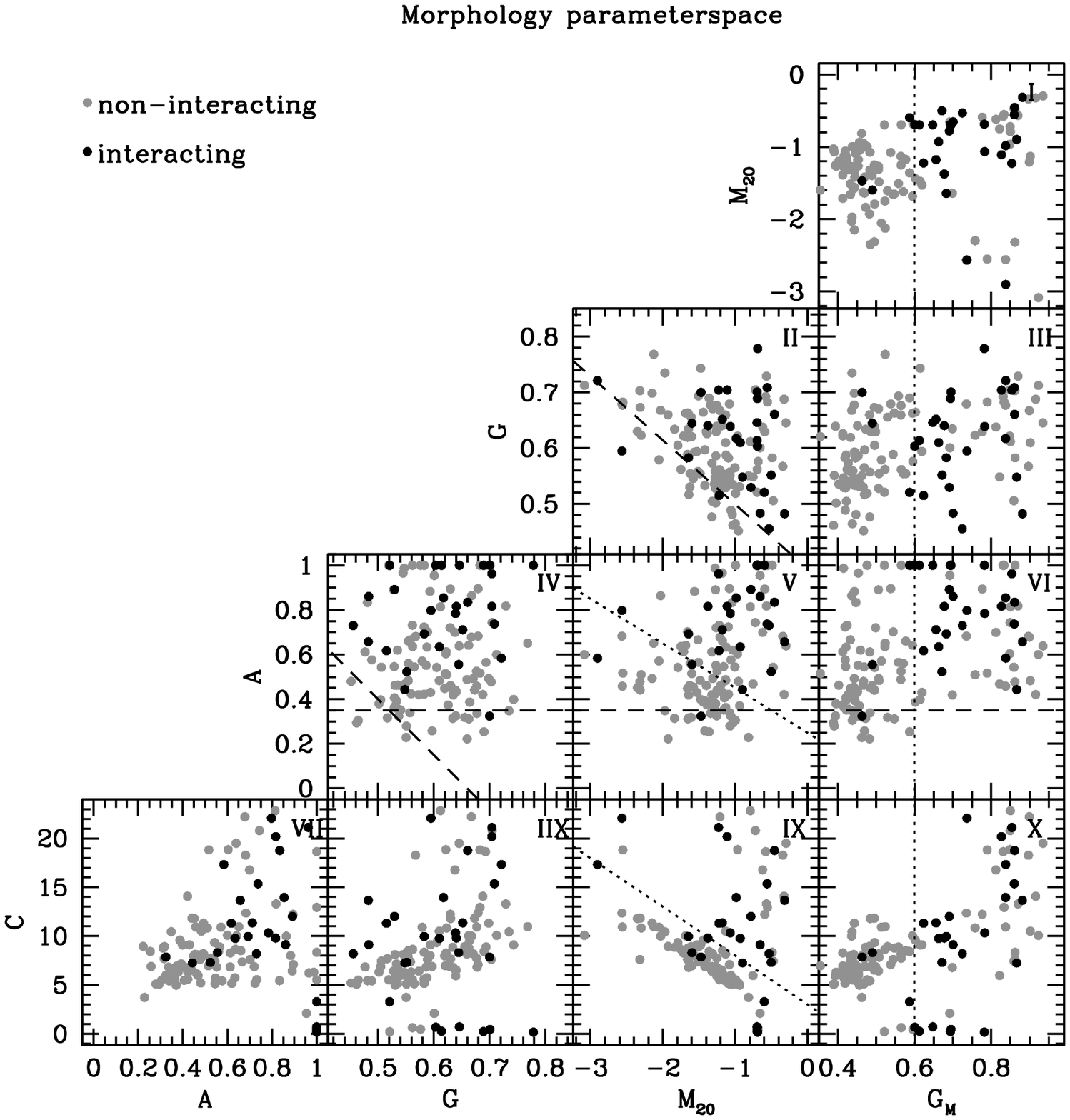}
\caption{\label{f:2d:spr:int} The parameter space of the combined samples from  \protect\cite{Swaters02} and \protect\cite{Noordermeer05} for interacting (black symbols) and non-interacting (gray symbols). There are clear separations in Asymmetry, $M_{20}$ and $G_M$ distributions between the two populations of galaxies. Our two cuts in parameters are indicated with a dotted line:    $A > 4.3 M_{20} $ and $G_M>0.6$. Combined, these two cuts select a reasonable fraction of the interacting galaxies. Values in Tables A1 and A2 {\em in the electronic version of the manuscript}. }
\end{figure*}

\subsection{Interaction}
\label{ss:int}

Based on the literature, one expects there be some signal of interaction in Concentration, Asymmetry, $M_{20}$ and Gini as the galaxy is warped and tidal arms are formed: the interaction spreads flux from the exponential distribution, altering Concentration and Gini, adds bright knots of stars further from the centre in optical images, changing both $M_{20}$ and Asymmetry, and interaction breaks the overall symmetry of the galaxy's image, also modifying Asymmetry. Since this parameter space was developed to classify galaxy morphology, one can expect changes in more than one of the parameters simultaneously when the spiral disk is gravitationally disturbed. During the interaction, one can also reasonably expect the disk to return periodically to unperturbed morphological values. Hence, we seek a section of this parameter space where interacting disks spend {\em some} of the several Gyr that an merger takes. The fraction of galaxies in this parameter space, together with and estimate of the typical time spent there gives a typical merger rate for an \hi \ survey.

Figure \ref{f:2d:hist:int} shows the histograms of the morphological parameters (Concentration, Asymmetry, Smoothness, Gini, $M_{20}$ and $G_M$), for interacting and non-interacting, isolated galaxies, according to \cite{Noordermeer05}. The five galaxies marked by \cite{Swaters02} as interacting are included. Again the interaction classification was done by a single observer, introducing some risk of a personal bias. The remainder of the Swaters et. al. sample is treated as isolated galaxies but it may contain some (mildly) interacting dwarf galaxies. In fact, given the observational result that lower mass galaxies at higher redshifts (z=0.2-1.2) show high fractions of interactions \citep[$\sim 10$\%][]{Bridge07, Bridge10, Kartaltepe07, Lotz08b, Lin08, Conselice09b, Jogee09, Lotz10a}, the real fraction of interacting galaxies may be higher. We therefore treat the Noordermeer sample as the cleanest and consider the Swaters sample more for confirmation.

Both the CAS space and the Gini/$M_{20}$ parameters have been used to identify morphologically disturbed galaxies in the literature and in the previous papers in this series, we established that the UV or FIR and the \hi \ perspective trace similar structure. Thus, we expect to see some signal of interaction in many of these parameters, notably Asymmetry, $M_{20}$ and $G_M$. 
The interacting galaxies show high values of Asymmetry ($A > 0.6$). This is somewhat higher than the cut used by \cite{Conselice06b} for optical asymmetry; $A_{optical} > 0.38$ but they required $A>S$ as well. The interacting galaxies have higher values of  $M_{20}$; $M_{20} ~ > ~ $-1, which is not too different from the cut used by \cite{Lotz04}. 
Direct cuts are most commonly used in these parameter spaces. Alternatively, one could define the eigenvectors of the interacting population in the combined parameter space. However, simply excluding the locus of non-interacting galaxies and including most of the merging ones is the best one can do since the interaction qualifier is a subjective and qualitative one, not a quantitative one like the tidal disturbance parameter used in \cite{Karachentsev04} and \cite{Bournaud05b}. \cite{Swaters02} and \cite{Noordermeer05} do not discern between weakly and strongly interacting. In addition, the training is likely too small to define eigenvectors. Therefore we use hard cuts in parameter space and compare how well these retrieve the fraction and objects that the Swaters and Noordermeer papers marked as interacting.

\begin{figure*}
\centering
\includegraphics[width=0.49\textwidth]{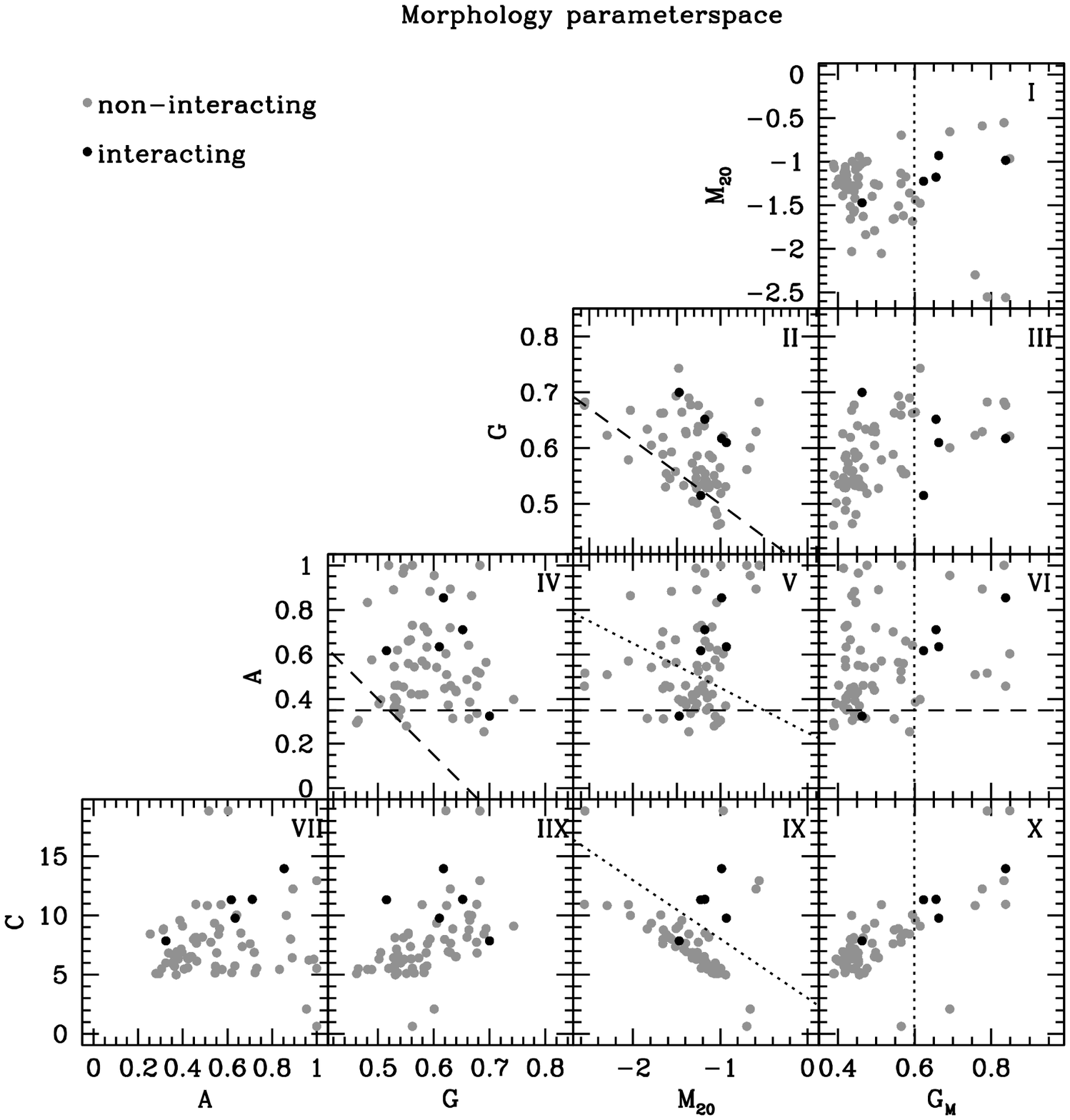}
\includegraphics[width=0.49\textwidth]{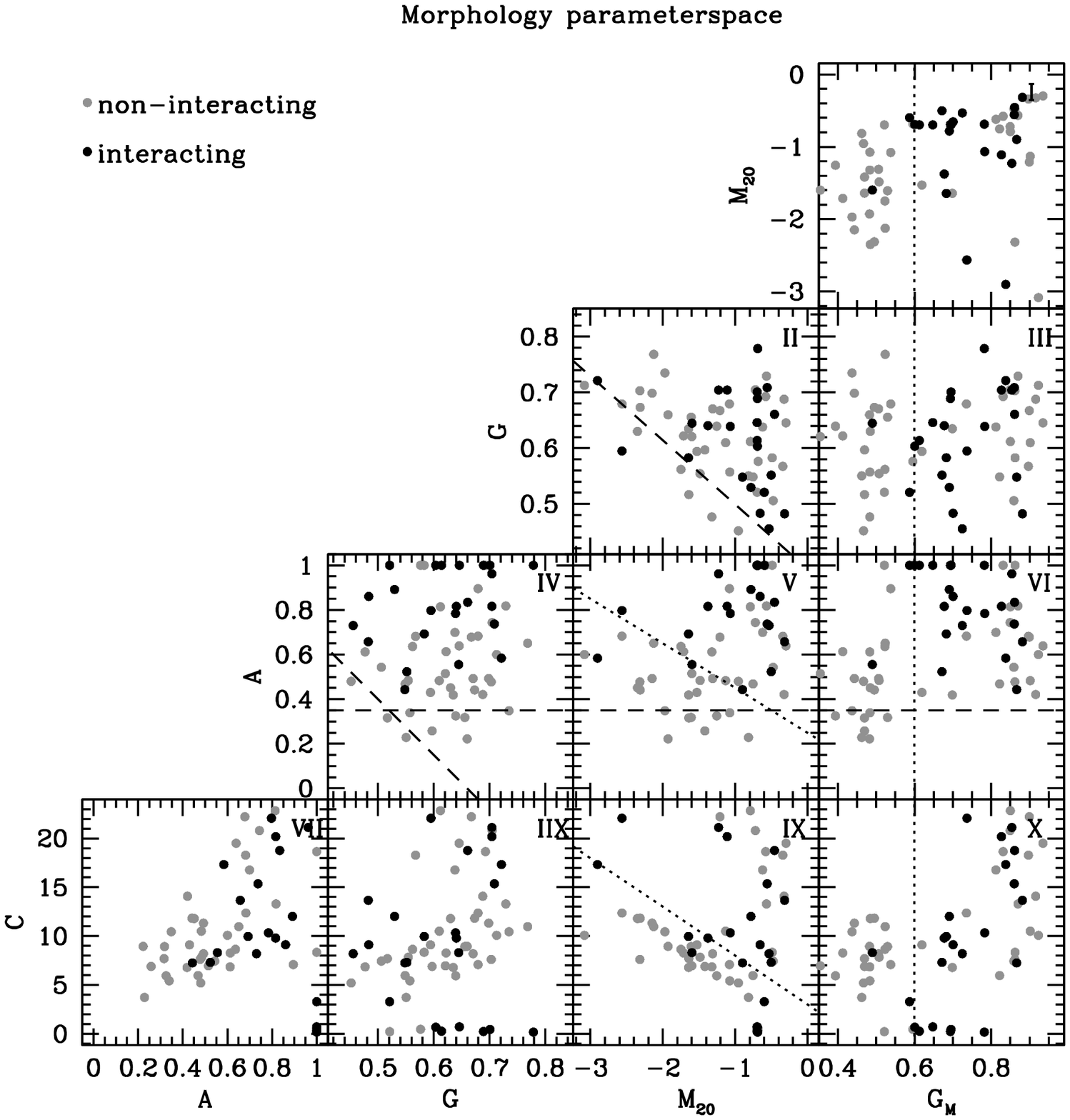}
\caption{\label{f:comp} The parameter space of the samples from  \protect\cite{Swaters02} (left) and \protect\cite{Noordermeer05} (right) for interacting (black dots) and non-interacting (gray dots). The Swaters et al sample is comprised of dwarf systems and the Noordermeer et al. one of bigger spirals. Values for the Swaters et al. (2002) and Noordermeer et al. (2005) sample can be found in Table A1 and A2 respectively {\em electronic version of the manuscript}.}
\end{figure*}

Following the example of \cite{Scarlata07}, we plot each morphological parameter against the others in Figure \ref{f:2d:spr:int} for the combined sample and for the Swaters and Noordermeer samples separately in Figure \ref{f:comp}. Both samples appear to occupy the same parameter space, so combining them is not an issue. Interacting galaxies are marked.
Figure \ref{f:2d:hist:int} confirms the assertion from \cite{CAS} and \cite{Lotz04} that Asymmetry and $M_{20}$ are parameters sensitive to mergers, and our assertion in \cite{Holwerda10a,Holwerda10b,Holwerdapra09} that \hi \ is a good wavelength to investigate it. 
We define three criteria to select mergers: 
\begin{equation}
G_M > 0.6,
\label{eq:crit1}
\end{equation}
\begin{equation}
A < -0.2 \times M_{20} + 0.25,
\label{eq:crit2}
\end{equation}
and
\begin{equation}
C > -5 \times M_{20} + 3. 
\label{eq:crit3}
\end{equation}
\noindent These are the dotted lines in Figures \ref{f:2d:spr:int} and \ref{f:comp} (equation \ref{eq:crit1} in Panels I, III, VI and X, equation \ref{eq:crit2} in panel V and equation \ref{eq:crit3} in panel IX). We also use three criteria for these parameters defined in the literature:
\begin{equation}
A> 0.38 
\label{eq:lcrit1}
\end{equation}
\begin{equation}
G > -0.115 \times M_{20} + 0.384
\label{eq:lcrit2}
\end{equation}
and
\begin{equation}
G > -0.4 \times A + 0.66 ~{\rm or} ~ A > 0.4
\label{eq:lcrit2}
\end{equation}
With the first one from \cite{CAS} and the last two from \cite{Lotz04,Lotz10b}. These are the dashed lines in Figures \ref{f:2d:spr:int} and \ref{f:comp}, panels IV-VI, II and IV respectively.

Combined with those from the literature, we list their success rates in Table \ref{t:frac}. From this table, it is evident that the criteria from the literature do not translate well to \hi \ column density map morphology. These criteria select too many contaminants. In part this may be because both Gini and Concentration are linked and sensitive to how concentrated the image is. As \hi \ maps are more extended, there is a shift in values \citep[see paper II,][]{Holwerda10b}. 
%
The $G_M$ parameter criterion performs well, selecting most interacting galaxies with $G_M > 0.6$ but with quite some contamination, even in the Noordermeer sample (Table \ref{t:frac}) which is the cleanest of the two. 
Therefore, a combination of one or more morphological parameters appears the most promising to cleanly separate an \hi \ sample into interacting and isolated galaxies.
The Asymetry-$M_{20}$ selection criterion performs better in that it selects a similar fraction of galaxies but it does not agree with either Swaters or Noordermeer estimate in the case of individual objects. It does select many more objects in the Swaters sample but as we have noted above, one can reasonable expect more dwarfs to be (mildly) interacting than the five flagged by \cite{Swaters02}. 
The Concentration-$M_{20}$ criterion works best as it selects a similar fraction of galaxies as the visual classification but also agrees on more cases of individual galaxies. It also agrees well with the Swaters selection of interactions. Alternatively, a Concentration-Asymmetry criterion may well work.
%
For instance, combining any two of these criteria does not improve the selection appreciably. Combining the $G_M$ criterion and the C/$M_{20}$ criterion, effectively is the latter criterion. 

We intend to apply these morphological cuts on representative samples of \hi \ observations, starting with the complete WHISP sample \citep{Holwerda10e}. To convert these fraction into a volume merger rate, one needs to compute the representative volume of the survey and a timescale for which merging systems reside in the interaction part of parameter space. We focus on these timescales in the next paper in this series \citep{Holwerda10d}, using SPH simulations of gas-rich 1:1 mergers.

\section{Conclusions}
\label{s:concl}

Based on our quantified morphological analysis of \nsamp galaxies from \cite{Swaters02} and \cite{Noordermeer05} for which they provided visual estimates of the  lopsidedness and level of interaction for \hi \ disks, we can conclude the following:

\begin{itemize}
\item[1.] The two-dimensional morphological parameters cannot discriminate between weak and strong lopsidedness as judged visually by previous authors. However, Asymmetry, $M_{20}$ and $G_M$, and to a lesser extent Gini parameters all show a shift in the mean of the distribution of values between the lopsided galaxies and those that are not lopsided (Figure \ref{f:2d:hist:mlob}).
\item[2.] We suggest, therefore, that these parameters can be used to assign a probability of lopsidedness (Figure \ref{f:2d:hist:mlob}).
But future surveys should use Fourier analysis to find lopsidedness in the \hi \ distribution.
\item[3.] The fraction of interactions in a sample of \hi \ maps can however be determined similarly well using these parameters, as a visual classification. Individual parameters, such as Asymmetry and $G_M$, do not select the interacting systems cleanly (Figure \ref{f:2d:hist:int} and \ref{f:2d:spr:int}). 
\item[4.] Combined criteria, using Asymmetry and $M_{20}$ or Concentration and $M_{20}$, work better (Figure \ref{f:2d:spr:int} and \ref{f:comp}, and Table \ref{t:frac}) and select the right fraction of a sample of galaxies is currently undergoing interaction, as identified by visual inspection. Combined with an estimate of the time a merger is selected by these criteria, one can estimate what the merger rate in an \hi \ survey is. The benefits of such a merger rate determination would be less observer bias than a visual classification and a empirical visibility time, determined from simulations.
\end{itemize}

\section{Future Applications}
\label{s:fut}

The parameter space, as we applied it to the WHISP \hi \ column density maps, allows us to find candidates for lopsidedness and more accurately define the fraction of interacting galaxies, solely from their \hi \ morphology. It remains to be determined how long an interacting disk remains in the interaction part of the morphology parameter space. This can be addressed with the new generation of  simulations of major and minor mergers currently being undertaken \citep[e.g,][]{Bournaud05b,Cox06a, Cox06b,Weniger09,Lotz10a,Lotz10b}, which include a comprehensive treatment of the interstellar matter in the galaxies during the merger. The time-scale for which a disk has a morphological interaction signature can then be determined by averaging over the many possible viewing angles. This timescale and the full WHISP sample (368 galaxies) will allow us to estimate the interaction rate of spirals locally, based purely on their \hi \ morphology. This can serve as an additional zero-point for estimates of the merger rate at higher redshift. Upcoming nearby galaxy surveys with MeerKAT and the Widefield ASKAP L-band Legacy All-sky Blind surveY (WALLABY, Koribalski et al. {\em in preparation}) for the Southern Sky and Northern Sky Survey with APERTIF on WSRT will solidify the local Universe merger rate estimate, based on \hi \ morphology. The future Square Kilometre Array can subsequently determine the merger rate of gas-rich galaxies over Cosmic times (up to z$\sim$1 or better). 
The great benefit of \hi \ surveys to determine the merger rates are the sensitivity of \hi \ to interaction and the sensitivity of \hi \ surveys to lower mass systems, for which the merger rate is the poorest constrained \citep[see][]{Lotz10a,Lotz10b}. 

\section*{Acknowledgements}

The authors would like to thank W.C. Clarkson and J. Lotz for useful discussions and the anonymous referee for his or her careful comments and suggestions.

The work of B.W. Holwerda  and W.J.G. de Blok is based upon research supported by the South African Research Chairs Initiative of the Department of Science and Technology and the National Research Foundation. A. Bouchard acknowledges the financial support from the South African Square Kilometre Array Project.

The WHISP observation were carried out with the Westerbork Synthesis Radio Telescope, which is operated by the Netherlands Foundation for Research in Astronomy (ASTRON) with financial support from the Netherlands Foundation for Scientific Research (NWO). The WHISP project was carried out at the Kapteyn Astronomical Institute by J. Kamphuis, D. Sijbring and Y. Tang under the supervision of T.S. van Albada, J.M. van der Hulst and R. Sancisi. The Westerbork on the Web project was done by P. Kamphuis.
In memory of Dr. M.J. Holwerda. 
%


\begin{thebibliography}{79}
\expandafter\ifx\csname natexlab\endcsname\relax\def\natexlab#1{#1}\fi

\bibitem[{{Abraham} {et~al.}(1994){Abraham}, {Valdes}, {Yee}, \& {van den
  Bergh}}]{Abraham94}
{Abraham} R.~G., {Valdes} F., {Yee} H.~K.~C., {van den Bergh} S., 1994, \apj,
  432, 75

\bibitem[{{Abraham} {et~al.}(2003){Abraham}, {van den Bergh}, \&
  {Nair}}]{Abraham03}
{Abraham} R.~G., {van den Bergh} S., {Nair} P., 2003, \apj, 588, 218

\bibitem[{{Angiras} {et~al.}(2007){Angiras}, {Jog}, {Dwarakanath}, \&
  {Verheijen}}]{Angiras07}
{Angiras} R.~A., {Jog} C.~J., {Dwarakanath} K.~S., {Verheijen} M.~A.~W., 2007,
  \mnras, 378, 276

\bibitem[{{Angiras} {et~al.}(2006){Angiras}, {Jog}, {Omar}, \&
  {Dwarakanath}}]{Angiras06}
{Angiras} R.~A., {Jog} C.~J., {Omar} A., {Dwarakanath} K.~S., 2006, \mnras,
  369, 1849

\bibitem[{{Baldwin} {et~al.}(1980){Baldwin}, {Lynden-Bell}, \&
  {Sancisi}}]{Baldwin80}
{Baldwin} J.~E., {Lynden-Bell} D., {Sancisi} R., 1980, \mnras, 193, 313

\bibitem[{{Bendo} {et~al.}(2007){Bendo}, {Calzetti}, {Engelbracht},
  {Kennicutt}, {Meyer}, {Thornley}, {Walter}, {Dale}, {Li}, \&
  {Murphy}}]{Bendo07}
{Bendo} G.~J., {Calzetti} D., {Engelbracht} C.~W., {Kennicutt} R.~C., {Meyer}
  M.~J., {Thornley} M.~D., {Walter} F., {Dale} D.~A., {Li} A., {Murphy} E.~J.,
  2007, \mnras, 380, 1313

\bibitem[{{Bershady} {et~al.}(2000){Bershady}, {Jangren}, \&
  {Conselice}}]{Bershady00}
{Bershady} M.~A., {Jangren} A., {Conselice} C.~J., 2000, \aj, 119, 2645

\bibitem[{{Bertin} \& {Arnouts}(1996)}]{se}
{Bertin} E., {Arnouts} S., 1996, \aaps, 117, 393, provided by the NASA
  Astrophysics Data System

\bibitem[{{Booth} {et~al.}(2009){Booth}, {de Blok}, {Jonas}, \&
  {Fanaroff}}]{MeerKAT}
{Booth} R.~S., {de Blok} W.~J.~G., {Jonas} J.~L., {Fanaroff} B., 2009, ArXiv
  e-prints/0910.2935

\bibitem[{{Bournaud} {et~al.}(2005{\natexlab{a}}){Bournaud}, {Combes}, {Jog},
  \& {Puerari}}]{Bournaud05}
{Bournaud} F., {Combes} F., {Jog} C.~J., {Puerari} I., 2005{\natexlab{a}},
  \aap, 438, 507

\bibitem[{{Bournaud} {et~al.}(2005{\natexlab{b}}){Bournaud}, {Jog}, \&
  {Combes}}]{Bournaud05b}
{Bournaud} F., {Jog} C.~J., {Combes} F., 2005{\natexlab{b}}, \aap, 437, 69

\bibitem[{{Bridge} {et~al.}(2007){Bridge}, {Appleton}, {Conselice}, {Choi},
  {Armus}, {Fadda}, {Laine}, {Marleau}, {Carlberg}, {Helou}, \&
  {Yan}}]{Bridge07}
{Bridge} C.~R., {Appleton} P.~N., {Conselice} C.~J., {Choi} P.~I., {Armus} L.,
  {Fadda} D., {Laine} S., {Marleau} F.~R., {Carlberg} R.~G., {Helou} G., {Yan}
  L., 2007, \apj, 659, 931

\bibitem[{{Bridge} {et~al.}(2010){Bridge}, {Carlberg}, \&
  {Sullivan}}]{Bridge10}
{Bridge} C.~R., {Carlberg} R.~G., {Sullivan} M., 2010, \apj, 709, 1067

\bibitem[{{Bundy} {et~al.}(2005){Bundy}, {Ellis}, \& {Conselice}}]{Bundy05}
{Bundy} K., {Ellis} R.~S., {Conselice} C.~J., 2005, \apj, 625, 621

\bibitem[{{Carilli} \& {Rawlings}(2004)}]{ska}
{Carilli} C.~L., {Rawlings} S., 2004, New Astronomy Review, 48, 979

\bibitem[{{Conselice}(2003)}]{CAS}
{Conselice} C.~J., 2003, \apjs, 147, 1

\bibitem[{{Conselice}(2006)}]{Conselice06b}
---, 2006, \apj, 638, 686

\bibitem[{{Conselice}(2009)}]{Conselice09c}
---, 2009, \mnras, 399, L16

\bibitem[{{Conselice} {et~al.}(2000){Conselice}, {Bershady}, \&
  {Jangren}}]{Conselice00a}
{Conselice} C.~J., {Bershady} M.~A., {Jangren} A., 2000, \apj, 529, 886

\bibitem[{{Conselice} {et~al.}(2008){Conselice}, {Bundy}, {U}, {Eisenhardt},
  {Lotz}, \& {Newman}}]{Conselice08b}
{Conselice} C.~J., {Bundy} K., {U} V., {Eisenhardt} P., {Lotz} J., {Newman} J.,
  2008, \mnras, 383, 1366

\bibitem[{{Conselice} {et~al.}(2009){Conselice}, {Yang}, \&
  {Bluck}}]{Conselice09b}
{Conselice} C.~J., {Yang} C., {Bluck} A.~F.~L., 2009, \mnras, 361

\bibitem[{{Cox} {et~al.}(2006{\natexlab{a}}){Cox}, {Dutta}, {Di Matteo},
  {Hernquist}, {Hopkins}, {Robertson}, \& {Springel}}]{Cox06a}
{Cox} T.~J., {Dutta} S.~N., {Di Matteo} T., {Hernquist} L., {Hopkins} P.~F.,
  {Robertson} B., {Springel} V., 2006{\natexlab{a}}, \apj, 650, 791

\bibitem[{{Cox} {et~al.}(2006{\natexlab{b}}){Cox}, {Jonsson}, {Primack}, \&
  {Somerville}}]{Cox06b}
{Cox} T.~J., {Jonsson} P., {Primack} J.~R., {Somerville} R.~S.,
  2006{\natexlab{b}}, \mnras, 373, 1013

\bibitem[{{de Blok} {et~al.}(2009){de Blok}, {Jonas}, {Fanaroff}, {Holwerda},
  {Bouchard}, {Blyth}, {van der Heyden}, \& {Pirzkal}}]{meerkat2}
{de Blok} W.~J.~G., {Jonas} J., {Fanaroff} B., {Holwerda} B.~W., {Bouchard} A.,
  {Blyth} S., {van der Heyden} K., {Pirzkal} N., 2009, in Conference
  Proceedings of the "Panoramic Radio Astronomy: Wide-field 1-2 GHz research on
  galaxy evolution", June 02 - 05, 2009 Groningen, the Netherlands

\bibitem[{{Dury} {et~al.}(2008){Dury}, {De Rijcke}, {Debattista}, \&
  {Dejonghe}}]{Dury08}
{Dury} V., {De Rijcke} S., {Debattista} V.~P., {Dejonghe} H., 2008, ArXiv
  e-prints, 803

\bibitem[{{Gini}(1912)}]{Gini12}
{Gini} C., 1912

\bibitem[{{Haynes} {et~al.}(1998){Haynes}, {van Zee}, {Hogg}, {Roberts}, \&
  {Maddalena}}]{Haynes98}
{Haynes} M.~P., {van Zee} L., {Hogg} D.~E., {Roberts} M.~S., {Maddalena} R.~J.,
  1998, \aj, 115, 62

\bibitem[{{Holwerda}(2005)}]{seman}
{Holwerda} B.~W., 2005, astro-ph/0512139

\bibitem[{{Holwerda} {et~al.}(2009){Holwerda}, {de Blok}, {Bouchard}, {Blyth},
  {van der Heyden}, \& {Pirzkal}}]{Holwerdapra09}
{Holwerda} B.~W., {de Blok} W.~J.~G., {Bouchard} A., {Blyth} S., {van der
  Heyden} K., {Pirzkal} N., 2009, in Conference Proceedings of the "Panoramic
  Radio Astronomy: Wide-field 1-2 GHz research on galaxy evolution", June 02 -
  05, 2009 Groningen, the Netherlands

\bibitem[{{Holwerda} {et~al.}(2011{\natexlab{a}}){Holwerda}, {Pirzkal}, {de
  Blok}, {Blyth}, {Bouchard}, \& {van der Heyden}}]{Holwerda10e}
{Holwerda} B.~W., {Pirzkal} N., {de Blok} W.~J.~G., {Blyth} S.-L., {Bouchard}
  A., {van der Heyden} K.~J., 2011{\natexlab{a}}, \mnras, 4, {\it submitted}

\bibitem[{{Holwerda} {et~al.}(2011{\natexlab{b}}){Holwerda}, {Pirzkal}, {de
  Blok}, {Blyth}, {Bouchard}, {van der Heyden}, \& {Elson}}]{Holwerda10a}
{Holwerda} B.~W., {Pirzkal} N., {de Blok} W.~J.~G., {Blyth} S.-L., {Bouchard}
  A., {van der Heyden} K.~J., {Elson} E.~C., 2011{\natexlab{b}}, \mnras, 0,
  {\it submitted}

\bibitem[{{Holwerda} {et~al.}(2011{\natexlab{c}}){Holwerda}, {Pirzkal}, {de
  Blok}, {Blyth}, {Bouchard}, {van der Heyden}, \& {Elson}}]{Holwerda10b}
---, 2011{\natexlab{c}}, \mnras, 1, {\it accepted}

\bibitem[{{Holwerda} {et~al.}(2011{\natexlab{d}}){Holwerda}, {Pirzkal}, {de
  Blok}, {Blyth}, {Bouchard}, {van der Heyden}, \& {Elson}}]{Holwerda10d}
---, 2011{\natexlab{d}}, \mnras, 3, {\it submitted}

\bibitem[{{Jog}(1997)}]{Jog97}
{Jog} C.~J., 1997, \apj, 488, 642

\bibitem[{{Jog}(1999)}]{Jog99}
---, 1999, \apj, 522, 661

\bibitem[{{Jog} \& {Combes}(2009)}]{Jog09}
{Jog} C.~J., {Combes} F., 2009, \physrep, 471, 75

\bibitem[{{Jogee} {et~al.}(2009){Jogee}, {Miller}, {Penner}, {Skelton},
  {Conselice}, {Somerville}, {Bell}, {Zheng}, {Rix}, {Robaina}, {Barazza},
  {Barden}, {Borch}, {Beckwith}, {Caldwell}, {Peng}, {Heymans}, {McIntosh},
  {H{\"a}u{\ss}ler}, {Jahnke}, {Meisenheimer}, {Sanchez}, {Wisotzki}, {Wolf},
  \& {Papovich}}]{Jogee09}
{Jogee} S., {Miller} S.~H., {Penner} K., {Skelton} R.~E., {Conselice} C.~J.,
  {Somerville} R.~S., {Bell} E.~F., {Zheng} X.~Z., {Rix} H., {Robaina} A.~R.,
  {Barazza} F.~D., {Barden} M., {Borch} A., {Beckwith} S.~V.~W., {Caldwell}
  J.~A.~R., {Peng} C.~Y., {Heymans} C., {McIntosh} D.~H., {H{\"a}u{\ss}ler} B.,
  {Jahnke} K., {Meisenheimer} K., {Sanchez} S.~F., {Wisotzki} L., {Wolf} C.,
  {Papovich} C., 2009, \apj, 697, 1971

\bibitem[{{Johnston}(2007)}]{askap2}
{Johnston} S., 2007, in From Planets to Dark Energy: the Modern Radio Universe.
  October 1-5 2007, The University of Manchester, UK. Published online at
  SISSA, Proceedings of Science, p.6

\bibitem[{{Johnston} {et~al.}(2007){Johnston}, {Bailes}, {Bartel}, {Baugh},
  {Bietenholz}, {Blake}, {Braun}, {Brown}, {Chatterjee}, {Darling}, {Deller},
  {Dodson}, {Edwards}, {Ekers}, {Ellingsen}, {Feain}, {Gaensler}, {Haverkorn},
  {Hobbs}, {Hopkins}, {Jackson}, {James}, {Joncas}, {Kaspi}, {Kilborn},
  {Koribalski}, {Kothes}, {Landecker}, {Lenc}, {Lovell}, {Macquart},
  {Manchester}, {Matthews}, {McClure-Griffiths}, {Norris}, {Pen}, {Phillips},
  {Power}, {Protheroe}, {Sadler}, {Schmidt}, {Stairs}, {Staveley-Smith},
  {Stil}, {Taylor}, {Tingay}, {Tzioumis}, {Walker}, {Wall}, \&
  {Wolleben}}]{askap1}
{Johnston} S., {Bailes} M., {Bartel} N., {Baugh} C., {Bietenholz} M., {Blake}
  C., {Braun} R., {Brown} J., {Chatterjee} S., {Darling} J., {Deller} A.,
  {Dodson} R., {Edwards} P.~G., {Ekers} R., {Ellingsen} S., {Feain} I.,
  {Gaensler} B.~M., {Haverkorn} M., {Hobbs} G., {Hopkins} A., {Jackson} C.,
  {James} C., {Joncas} G., {Kaspi} V., {Kilborn} V., {Koribalski} B., {Kothes}
  R., {Landecker} T.~L., {Lenc} E., {Lovell} J., {Macquart} J.-P., {Manchester}
  R., {Matthews} D., {McClure-Griffiths} N.~M., {Norris} R., {Pen} U.-L.,
  {Phillips} C., {Power} C., {Protheroe} R., {Sadler} E., {Schmidt} B.,
  {Stairs} I., {Staveley-Smith} L., {Stil} J., {Taylor} R., {Tingay} S.,
  {Tzioumis} A., {Walker} M., {Wall} J., {Wolleben} M., 2007, Publications of
  the Astronomical Society of Australia, 24, 174

\bibitem[{{Johnston} {et~al.}(2009){Johnston}, {Feain}, \& {Gupta}}]{askap4}
{Johnston} S., {Feain} I.~J., {Gupta} N., 2009, in Astronomical Society of the
  Pacific Conference Series, Vol. 407, Astronomical Society of the Pacific
  Conference Series, {D.~J.~Saikia, D.~A.~Green, Y.~Gupta, \& T.~Venturi}, ed.,
  pp. 446--+

\bibitem[{{Johnston} {et~al.}(2008{\natexlab{a}}){Johnston}, {Taylor},
  {Bailes}, {Bartel}, {Baugh}, {Bietenholz}, {Blake}, {Braun}, {Brown},
  {Chatterjee}, {Darling}, {Deller}, {Dodson}, {Edwards}, {Ekers}, {Ellingsen},
  {Feain}, {Gaensler}, {Haverkorn}, {Hobbs}, {Hopkins}, {Jackson}, {James},
  {Joncas}, {Kaspi}, {Kilborn}, {Koribalski}, {Kothes}, {Landecker}, {Lenc},
  {Lovell}, {Macquart}, {Manchester}, {Matthews}, {McClure-Griffiths},
  {Norris}, {Pen}, {Phillips}, {Power}, {Protheroe}, {Sadler}, {Schmidt},
  {Stairs}, {Staveley-Smith}, {Stil}, {Tingay}, {Tzioumis}, {Walker}, {Wall},
  \& {Wolleben}}]{ASKAP}
{Johnston} S., {Taylor} R., {Bailes} M., {Bartel} N., {Baugh} C., {Bietenholz}
  M., {Blake} C., {Braun} R., {Brown} J., {Chatterjee} S., {Darling} J.,
  {Deller} A., {Dodson} R., {Edwards} P., {Ekers} R., {Ellingsen} S., {Feain}
  I., {Gaensler} B., {Haverkorn} M., {Hobbs} G., {Hopkins} A., {Jackson} C.,
  {James} C., {Joncas} G., {Kaspi} V., {Kilborn} V., {Koribalski} B., {Kothes}
  R., {Landecker} T., {Lenc} A., {Lovell} J., {Macquart} J.-P., {Manchester}
  R., {Matthews} D., {McClure-Griffiths} N., {Norris} R., {Pen} U.-L.,
  {Phillips} C., {Power} C., {Protheroe} R., {Sadler} E., {Schmidt} B.,
  {Stairs} I., {Staveley-Smith} L., {Stil} J., {Tingay} S., {Tzioumis} A.,
  {Walker} M., {Wall} J., {Wolleben} M., 2008{\natexlab{a}}, Experimental
  Astronomy, 22, 151

\bibitem[{{Johnston} {et~al.}(2008{\natexlab{b}}){Johnston}, {Taylor},
  {Bailes}, {Bartel}, {Baugh}, {Bietenholz}, {Blake}, {Braun}, {Brown},
  {Chatterjee}, {Darling}, {Deller}, {Dodson}, {Edwards}, {Ekers}, {Ellingsen},
  {Feain}, {Gaensler}, {Haverkorn}, {Hobbs}, {Hopkins}, {Jackson}, {James},
  {Joncas}, {Kaspi}, {Kilborn}, {Koribalski}, {Kothes}, {Landecker}, {Lenc},
  {Lovell}, {Macquart}, {Manchester}, {Matthews}, {McClure-Griffiths},
  {Norris}, {Pen}, {Phillips}, {Power}, {Protheroe}, {Sadler}, {Schmidt},
  {Stairs}, {Staveley-Smith}, {Stil}, {Tingay}, {Tzioumis}, {Walker}, {Wall},
  \& {Wolleben}}]{askap3}
---, 2008{\natexlab{b}}, Experimental Astronomy, 22, 151

\bibitem[{{Jonas}(2007)}]{meerkat1}
{Jonas} J., 2007, in From Planets to Dark Energy: the Modern Radio Universe.
  October 1-5 2007, The University of Manchester, UK. Published online at
  SISSA, Proceedings of Science, p.7

\bibitem[{{Karachentsev} {et~al.}(2004){Karachentsev}, {Karachentseva},
  {Huchtmeier}, \& {Makarov}}]{Karachentsev04}
{Karachentsev} I.~D., {Karachentseva} V.~E., {Huchtmeier} W.~K., {Makarov}
  D.~I., 2004, \aj, 127, 2031

\bibitem[{{Kartaltepe} {et~al.}(2007){Kartaltepe}, {Sanders}, {Scoville},
  {Calzetti}, {Capak}, {Koekemoer}, {Mobasher}, {Murayama}, {Salvato},
  {Sasaki}, \& {Taniguchi}}]{Kartaltepe07}
{Kartaltepe} J.~S., {Sanders} D.~B., {Scoville} N.~Z., {Calzetti} D., {Capak}
  P., {Koekemoer} A., {Mobasher} B., {Murayama} T., {Salvato} M., {Sasaki}
  S.~S., {Taniguchi} Y., 2007, \apjs, 172, 320

\bibitem[{{Levine} \& {Sparke}(1998)}]{Levine98}
{Levine} S.~E., {Sparke} L.~S., 1998, \apjl, 496, L13+

\bibitem[{{Lin} {et~al.}(2008){Lin}, {Patton}, {Koo}, {Casteels}, {Conselice},
  {Faber}, {Lotz}, {Willmer}, {Hsieh}, {Chiueh}, {Newman}, {Novak}, {Weiner},
  \& {Cooper}}]{Lin08}
{Lin} L., {Patton} D.~R., {Koo} D.~C., {Casteels} K., {Conselice} C.~J.,
  {Faber} S.~M., {Lotz} J., {Willmer} C.~N.~A., {Hsieh} B.~C., {Chiueh} T.,
  {Newman} J.~A., {Novak} G.~S., {Weiner} B.~J., {Cooper} M.~C., 2008, \apj,
  681, 232

\bibitem[{{Lotz} {et~al.}(2008{\natexlab{a}}){Lotz}, {Davis}, {Faber},
  {Guhathakurta}, {Gwyn}, {Huang}, {Koo}, {Le Floc'h}, {Lin}, {Newman},
  {Noeske}, {Papovich}, {Willmer}, {Coil}, {Conselice}, {Cooper}, {Hopkins},
  {Metevier}, {Primack}, {Rieke}, \& {Weiner}}]{Lotz08b}
{Lotz} J.~M., {Davis} M., {Faber} S.~M., {Guhathakurta} P., {Gwyn} S., {Huang}
  J., {Koo} D.~C., {Le Floc'h} E., {Lin} L., {Newman} J., {Noeske} K.,
  {Papovich} C., {Willmer} C.~N.~A., {Coil} A., {Conselice} C.~J., {Cooper} M.,
  {Hopkins} A.~M., {Metevier} A., {Primack} J., {Rieke} G., {Weiner} B.~J.,
  2008{\natexlab{a}}, \apj, 672, 177

\bibitem[{{Lotz} {et~al.}(2008{\natexlab{b}}){Lotz}, {Jonsson}, {Cox}, \&
  {Primack}}]{Lotz08}
{Lotz} J.~M., {Jonsson} P., {Cox} T.~J., {Primack} J.~R., 2008{\natexlab{b}},
  \mnras, 391, 1137

\bibitem[{{Lotz} {et~al.}(2008{\natexlab{c}}){Lotz}, {Jonsson}, {Cox}, \&
  {Primack}}]{Lotz08a}
---, 2008{\natexlab{c}}, ArXiv e-prints, 805

\bibitem[{{Lotz} {et~al.}(2010{\natexlab{a}}){Lotz}, {Jonsson}, {Cox}, \&
  {Primack}}]{Lotz10a}
---, 2010{\natexlab{a}}, \mnras, 404, 590

\bibitem[{{Lotz} {et~al.}(2010{\natexlab{b}}){Lotz}, {Jonsson}, {Cox}, \&
  {Primack}}]{Lotz10b}
---, 2010{\natexlab{b}}, \mnras, 404, 575

\bibitem[{{Lotz} {et~al.}(2004){Lotz}, {Primack}, \& {Madau}}]{Lotz04}
{Lotz} J.~M., {Primack} J., {Madau} P., 2004, \aj, 128, 163

\bibitem[{{Lovelace} {et~al.}(1999){Lovelace}, {Zhang}, {Kornreich}, \&
  {Haynes}}]{Lovelace99}
{Lovelace} R.~V.~E., {Zhang} L., {Kornreich} D.~A., {Haynes} M.~P., 1999, \apj,
  524, 634

\bibitem[{{Malin}(1978)}]{Malin78b}
{Malin} D.~F., 1978, \nat, 276, 591

\bibitem[{{Mapelli} {et~al.}(2009){Mapelli}, {Moore}, \&
  {Bland-Hawthorn}}]{Mapelli09}
{Mapelli} M., {Moore} B., {Bland-Hawthorn} J., 2009, in IAU Symposium, Vol.
  254, IAU Symposium, {Andersen} J., {Bland-Hawthorn} J., {Nordstr{\"o}m} B.,
  eds., pp. 40P--+

\bibitem[{{Matthews} {et~al.}(1998){Matthews}, {van Driel}, \&
  {Gallagher}}]{Matthews98}
{Matthews} L.~D., {van Driel} W., {Gallagher} III J.~S., 1998, \aj, 116, 1169

\bibitem[{{Mu{\~n}oz-Mateos} {et~al.}(2009){Mu{\~n}oz-Mateos}, {Gil de Paz},
  {Zamorano}, {Boissier}, {Dale}, {P{\'e}rez-Gonz{\'a}lez}, {Gallego},
  {Madore}, {Bendo}, {Boselli}, {Buat}, {Calzetti}, {Moustakas}, \&
  {Kennicutt}}]{Munoz-Mateos09}
{Mu{\~n}oz-Mateos} J.~C., {Gil de Paz} A., {Zamorano} J., {Boissier} S., {Dale}
  D.~A., {P{\'e}rez-Gonz{\'a}lez} P.~G., {Gallego} J., {Madore} B.~F., {Bendo}
  G., {Boselli} A., {Buat} V., {Calzetti} D., {Moustakas} J., {Kennicutt}
  R.~C., 2009, \apj, 703, 1569

\bibitem[{{Nilson}(1973)}]{Nilson73}
{Nilson} P., 1973, {Uppsala general catalogue of galaxies}. Acta Universitatis
  Upsaliensis.~Nova Acta Regiae Societatis Scientiarum Upsaliensis - Uppsala
  Astronomiska Observatoriums Annaler, Uppsala: Astronomiska Observatorium,
  1973

\bibitem[{{Noordermeer} {et~al.}(2001){Noordermeer}, {Sparke}, \&
  {Levine}}]{Noordermeer01}
{Noordermeer} E., {Sparke} L.~S., {Levine} S.~E., 2001, \mnras, 328, 1064

\bibitem[{{Noordermeer} {et~al.}(2005{\natexlab{a}}){Noordermeer}, {van der
  Hulst}, {Sancisi}, {Swaters}, \& {van Albada}}]{Noordermeer05cat}
{Noordermeer} E., {van der Hulst} J.~M., {Sancisi} R., {Swaters} R.~A., {van
  Albada} T.~S., 2005{\natexlab{a}}, VizieR Online Data Catalog, 344, 20137

\bibitem[{{Noordermeer} {et~al.}(2005{\natexlab{b}}){Noordermeer}, {van der
  Hulst}, {Sancisi}, {Swaters}, \& {van Albada}}]{Noordermeer05}
---, 2005{\natexlab{b}}, \aap, 442, 137

\bibitem[{{Pirzkal} {et~al.}(2006){Pirzkal}, {Xu}, {Ferreras}, {Malhotra},
  {Mobasher}, {Rhoads}, {Pasquali}, {Panagia}, {Koekemoer}, {Ferguson}, \&
  {Gronwall}}]{Pirzkal06}
{Pirzkal} N., {Xu} C., {Ferreras} I., {Malhotra} S., {Mobasher} B., {Rhoads}
  J.~E., {Pasquali} A., {Panagia} N., {Koekemoer} A.~M., {Ferguson} H.~C.,
  {Gronwall} C., 2006, \apj, 636, 582

\bibitem[{{Ravindranath} {et~al.}(2006){Ravindranath}, {Giavalisco},
  {Ferguson}, {Conselice}, {Katz}, {Weinberg}, {Lotz}, {Dickinson}, {Fall},
  {Mobasher}, \& {Papovich}}]{Ravindranath06}
{Ravindranath} S., {Giavalisco} M., {Ferguson} H.~C., {Conselice} C., {Katz}
  N., {Weinberg} M., {Lotz} J., {Dickinson} M., {Fall} S.~M., {Mobasher} B.,
  {Papovich} C., 2006, \apj, 652, 963

\bibitem[{{Richter} \& {Sancisi}(1994)}]{Richter94}
{Richter} O.-G., {Sancisi} R., 1994, \aap, 290, L9

\bibitem[{{Rix} \& {Zaritsky}(1995)}]{Rix95}
{Rix} H.-W., {Zaritsky} D., 1995, \apj, 447, 82

\bibitem[{{Scarlata} {et~al.}(2007){Scarlata}, {Carollo}, {Lilly}, {Sargent},
  {Feldmann}, {Kampczyk}, {Porciani}, {Koekemoer}, {Scoville}, {Kneib},
  {Leauthaud}, {Massey}, {Rhodes}, {Tasca}, {Capak}, {Maier}, {McCracken},
  {Mobasher}, {Renzini}, {Taniguchi}, {Thompson}, {Sheth}, {Ajiki}, {Aussel},
  {Murayama}, {Sanders}, {Sasaki}, {Shioya}, \& {Takahashi}}]{Scarlata07}
{Scarlata} C., {Carollo} C.~M., {Lilly} S., {Sargent} M.~T., {Feldmann} R.,
  {Kampczyk} P., {Porciani} C., {Koekemoer} A., {Scoville} N., {Kneib} J.-P.,
  {Leauthaud} A., {Massey} R., {Rhodes} J., {Tasca} L., {Capak} P., {Maier} C.,
  {McCracken} H.~J., {Mobasher} B., {Renzini} A., {Taniguchi} Y., {Thompson}
  D., {Sheth} K., {Ajiki} M., {Aussel} H., {Murayama} T., {Sanders} D.~B.,
  {Sasaki} S., {Shioya} Y., {Takahashi} M., 2007, \apjs, 172, 406

\bibitem[{{Schoenmakers} {et~al.}(1997){Schoenmakers}, {Franx}, \& {de
  Zeeuw}}]{Schoenmakers97}
{Schoenmakers} R.~H.~M., {Franx} M., {de Zeeuw} P.~T., 1997, \mnras, 292, 349

\bibitem[{{Swaters} \& {Balcells}(2002)}]{Swaters02b}
{Swaters} R.~A., {Balcells} M., 2002, \aap, 390, 863

\bibitem[{{Swaters} {et~al.}(1999){Swaters}, {Schoenmakers}, {Sancisi}, \& {van
  Albada}}]{Swaters99}
{Swaters} R.~A., {Schoenmakers} R.~H.~M., {Sancisi} R., {van Albada} T.~S.,
  1999, \mnras, 304, 330

\bibitem[{{Swaters} {et~al.}(2002){Swaters}, {van Albada}, {van der Hulst}, \&
  {Sancisi}}]{Swaters02}
{Swaters} R.~A., {van Albada} T.~S., {van der Hulst} J.~M., {Sancisi} R., 2002,
  \aap, 390, 829

\bibitem[{{Takamiya}(1999)}]{Takamiya99}
{Takamiya} M., 1999, \apjs, 122, 109

\bibitem[{{Trachternach} {et~al.}(2008){Trachternach}, {de Blok}, {Walter},
  {Brinks}, \& {Kennicutt}}]{Trachternach08}
{Trachternach} C., {de Blok} W.~J.~G., {Walter} F., {Brinks} E., {Kennicutt}
  R.~C., 2008, \aj, 136, 2720

\bibitem[{{van der Hulst}(2002)}]{whisp2}
{van der Hulst} J.~M., 2002, in Astronomical Society of the Pacific Conference
  Series, Vol. 276, Seeing Through the Dust: The Detection of HI and the
  Exploration of the ISM in Galaxies, {Taylor} A.~R., {Landecker} T.~L.,
  {Willis} A.~G., eds., pp. 84--+

\bibitem[{{van der Hulst} {et~al.}(2001){van der Hulst}, {van Albada}, \&
  {Sancisi}}]{whisp}
{van der Hulst} J.~M., {van Albada} T.~S., {Sancisi} R., 2001, in Astronomical
  Society of the Pacific Conference Series, Vol. 240, Gas and Galaxy Evolution,
  {Hibbard} J.~E., {Rupen} M., {van Gorkom} J.~H., eds., pp. 451--+

\bibitem[{{van Eymeren} {et~al.}(2011){van Eymeren}, {Juette}, {Jog}, {Stein},
  \& {Dettmar}}]{van-Eymeren11b}
{van Eymeren} J., {Juette} E., {Jog} C.~J., {Stein} Y., {Dettmar} R., 2011,
  ArXiv e-prints

\bibitem[{{Weniger} {et~al.}(2009){Weniger}, {Theis}, \& {Harfst}}]{Weniger09}
{Weniger} J., {Theis} C., {Harfst} S., 2009, ArXiv e-prints

\bibitem[{{Yitzhaki}(1991)}]{Yitzhaki91}
{Yitzhaki} S., 1991, American Statistical Association, 9, 235

\bibitem[{{Zaritsky} \& {Rix}(1997)}]{Zaritsky97}
{Zaritsky} D., {Rix} H.-W., 1997, \apj, 477, 118

\end{thebibliography}

\end{document}